\documentclass[11pt]{article}
\usepackage{fullpage}
\usepackage{graphicx}
\usepackage{latexsym,amsmath,amssymb,amsthm,epic,eepic,multirow}
\usepackage{natbib}

\usepackage{multirow}
\usepackage{subfigure}
\usepackage{natbib}
\usepackage{algorithm}
\usepackage{algorithmic}

\newcommand{\PP}{\mathbb{P}}
\newcommand{\EE}{\mathbb{E}}
\newcommand{\R}{\mathbb{R}}

\newcommand{\eps}{\varepsilon}

\newcommand{\Cov}{\mbox{Cov}}

\newcommand{\bx}{\mathbf{X}}
\newcommand{\argmin}{\mathrm{argmin}}

\newtheorem{theo}{Theorem}
\newtheorem{prop}{Proposition}
\newtheorem{lemm}{Lemma}
\newtheorem{corr}{Corollary}

\usepackage{color}

\begin{document}

\title{High-dimensional inference in misspecified linear models}   
  
\author{Peter B\"uhlmann and Sara van de Geer}


\maketitle

\begin{abstract}
We consider high-dimensional inference when the assumed linear model is
misspecified. We describe some correct interpretations and corresponding
sufficient assumptions for valid asymptotic inference of the model
parameters, which still have a useful meaning when the model is 
misspecified. We largely focus on the de-sparsified Lasso procedure but we
also indicate some implications for (multiple) sample splitting
techniques. In view of available methods and software, our results
contribute to robustness considerations with respect to model
misspecification. 
\end{abstract}

\section{Introduction}

The construction of confidence intervals and statistical hypothesis tests
is a primary goal for assessing uncertainty in high-dimensional
inference. Most of the  recent contributions for this task discuss some
methods and  
approaches for high-dimensional linear models
\citep{pb13,zhangzhang11,vdgetal13,jamo13b,meins13,foygcand14},
but generalized linear models \citep{memepb09,minnieretal11,vdgetal13},
undirected graphical models \citep{renetal13,jankvdg14}, instrumental
variable models \citep{belloni2012sparse} or very general models
\citep{mebu10} have been considered as well, and all of these latter
references cover linear models as special case. Another philosophy for
inference in the high-dimensional setting is based on selective inference
\citep{beye05,covtest14,tayetal14}, but we do not consider this here. Our
goal is to interpret and analyze the meaning of inference 
procedures when the linear model is misspecified. We address this issue in
greater detail for the de-sparsified (or de-biased) Lasso
\citep{zhangzhang11}, but we make a few more general comments in Section
\ref{subsec.samplesplit}.

More concretely, we describe the correct interpretations and corresponding
(sufficient) assumptions which guarantee valid asymptotic inference for
the parameters in a high-dimensional, misspecified linear model. That is, we
assume that the data is generated from an 
underlying true nonlinear model $Y = f(X) + \xi$ but we fit the wrong
linear model $Y = X \beta^0 + \eps$ to the data; see for example
\citet{wass14} who describes such settings as ``weak modeling''. Precise
definitions of the 
models are given later. Some arising questions are: first, what is the
interpretation of $\beta^0$; and secondly, is the standard de-sparsified
Lasso procedure valid for 
construction of statistical hypothesis tests and confidence intervals for
the components $\beta^0_j\ (j=1,\ldots ,p)$.
Regarding the first issue, it is important to
distinguish between random and fixed design scenarios. Regarding the
second point, we do give sufficient conditions for asymptotic correctness of the
de-sparsified Lasso procedure, although for the random design case, one has
to estimate the asymptotic variance differently than for correctly
specified models.  

The novelty of this work is that we explicitly discuss the implications of
linear model misspecification for construction of confidence intervals and
hypothesis testing in high dimensions. We believe that this is a missing piece
which should be addressed 
and which is informally often treated according to the folklore that the
procedure leads to inference for the 
``best projected regression parameters'': we make this 
precise and also show that some modifications are necessary for the random
design case (see above). The latter are implemented in the statistical
\texttt{R}-software package \texttt{hdi} \citep{hdipackage} which includes
various methods for frequentist high-dimensional inference
\citep{dezeureetal14}.   

\section{The de-sparsified Lasso for potentially misspecified linear models}

We consider $n$ data points $(Y^{(1)},X^{(1)}),\ldots,(Y^{(n)},X^{(n)})$
with univariate responses $Y^{(i)}$ and $p$-dimensional covariables
$X^{(i)}$. We denote by 
$Y = (Y^{(1)},\ldots ,Y^{(n)})^T$ and $X_j = (X_j^{(1)},\ldots
,X_j^{(n)})^T\ (j=1,\ldots ,p)$ the $n \times 1$ vectors, and by $\bx =
(X_1,\ldots ,X_p)$ the $n \times p$ design matrix.  

We fit a potentially misspecified linear model 
\begin{eqnarray}\label{lin.mod}
Y = \bx \beta^0 + \eps,
\end{eqnarray}
where the model assumptions are as follows: i.i.d. distributed rows of $\bx$ (if
random), and i.i.d. components of $\eps$ having mean zero, variance
$\sigma_{\eps}^2$ and which are uncorrelated from $\bx$. In a misspecified
setting, the meaning of the parameter vector $\beta^0$ and of the errors
$\eps$ depends on the context, in particular whether the design is random
or fixed. The different interpretations are presented in Sections
\ref{sec.randomdes} and \ref{sec.fixeddesign} below. 

For constructing confidence intervals and hypothesis tests for the
individual parameters $\beta^0_j\ (j = 1,\ldots ,p)$, we consider
the de-sparsified Lasso, originally proposed by \citet{zhangzhang11}. The
procedure is as follows. First, do a Lasso \citep{tibs96} or square root
Lasso \citep{belloni2011square} regression fit of $X_j$ versus all other 
variables from $\bx_{-j}$, the $n \times (p-1)$ design matrix whose columns
correspond to the variables $\{X_k;\ k \neq j\}$. That is, for the Lasso,
\begin{eqnarray}\label{nodewiselasso}
\hat{\gamma}_j = \argmin_{\gamma \in \R^{p-1}}\left(\|X_j - \bx_{-j}
  \gamma\|_2^2/n + \lambda_X\|\gamma\|_1\right).
\end{eqnarray}
or using the square root Lasso,
\begin{eqnarray}\label{nodewisesqrtlasso}
\hat{\gamma}_j = \argmin_{\gamma \in \R^{p-1}}\left(\|X_j - \bx_{-j}
  \gamma\|_2/\sqrt{n} + \lambda_X\|\gamma\|_1\right).
\end{eqnarray}
The residuals of such a regression are denoted by
\begin{eqnarray*}
Z_j = X_j - \bx_{-j} \hat{\gamma}_j.
\end{eqnarray*}
We then project the response $Y$ onto this residual vector: if the model
(\ref{lin.mod}) were correct, we have
\begin{eqnarray*}
\frac{Z_j^T Y}{Z_j^T X_j} = \beta^0_j + \sum_{k\neq j} \frac{Z_j^T
  \bx_k}{Z_j^T X_j}\beta^0_k + \frac{Z_j^T \eps}{Z_j^T X_j}.
\end{eqnarray*}
This suggests a bias correction as follows. Pursue a Lasso regression of $Y$
versus $\bx$:
\begin{eqnarray*}
\hat{\beta} = \argmin_{\beta} \left(\|Y - \bx \beta\|_2^2/n + \lambda
\|\beta\|_1 \right),
\end{eqnarray*}
plug it into the bias term and subtract the estimated bias. This leads
to the de-sparsified Lasso estimator: 
\begin{eqnarray}\label{desplasso}
\hat{b}_j = \frac{Z_j^T Y}{Z_j^T X_j} - \sum_{k\neq j} \frac{Z_j^T
  \bx_k}{Z_j^T X_j}\hat{\beta}_k\ \ (j=1,\ldots ,p). 
\end{eqnarray}

From the construction and assuming that model (\ref{lin.mod}) is correct, we
heuristically obtain:  
\begin{eqnarray*}
\frac{Z_j^T X_j}{\sqrt{n} \omega_{p;jj}} (\hat{b}_j - \beta^0_j) = \sum_{k
  \neq j} \frac{Z_j^T  \bx_k}{\sqrt{n} \omega_{p;jj}}(\hat{\beta}_k -
\beta^0_k) + \frac{Z_j^T 
  \eps}{\sqrt{n} \omega_{p;jj}} \approx \frac{Z_j^T
  \eps}{\sqrt{n}\omega_{p;jj}} \approx {\cal N}(0, 1), 
\end{eqnarray*}
where we assume for the first approximation that the error in estimating
the bias is negligible, and where $\omega_{p;jj}^2$ is the asymptotic
variance of $Z_j^T \eps/\sqrt{n}$. This
reasoning has been made rigorous in earlier work, assuming some conditions
\citep{zhangzhang11,vdgetal13}. 
When the model (\ref{lin.mod}) is wrong, however, the heuristics above
needs to be justified anew. Also from a practical point of view,
we need to characterize the meaning for $\beta^0$ and we need to determine
the correct specification of $\omega_{p;jj}^2$ in order to construct
asymptotically correct confidence intervals and tests. 
The details are described in the following Sections
\ref{sec.randomdes} and \ref{sec.fixeddesign}. 

The procedure for the de-sparsified Lasso $\hat{b}_j$ in (\ref{desplasso})
remains (essentially) the same regardless whether the linear model is correct or
not. Referring to the parenthesis in the previous sentence, what
potentially changes
relative to a correctly specified model is the proper asymptotic variance
$\omega_{p;jj}^2$, see Section \ref{subsec.estvar}, and this new feature is
now also implemented in the \textrm{R}-software package
\texttt{hdi} \citep{hdipackage}. 

Throughout the
paper, the asymptotic statements are for the setting where the
dimension $p = p_n$ is allowed to depend on $n$ (and hence also the random
variables in the model), and we consider the behavior as $n \to \infty$,
typically with $p = p_n \to \infty$ at a much faster rate than
$n$. We often suppress the index $n$ in the notation. 

\section{Random design model}\label{sec.randomdes}
Consider the true model
\begin{eqnarray}\label{mod.true}
Y^{(0)} = f^0(X^{(0)}) + \xi^{(0)},
\end{eqnarray}
where $\xi^{(0)}$ is independent of $ X^{(0)}$ with $\EE[\xi^{(0)}] = 0$.
For simplicity, we assume that $\EE[f^0(X^{(0)})] = 0$ as well
as $\EE[X^{(0)}] = 0$, and that furthermore the second moments of
$X^{(0)}$ and $Y^{(0)}$ exist. We assume that the data are realizations of
$(Y^{(1)},X^{(1)}),\ldots ,(Y^{(n)},X^{(n)})$ of i.i.d. copies of
$(Y^{(0)},X^{(0)})$ from model 
(\ref{mod.true}).  

Consider the linear projection
\begin{eqnarray}\label{betarand}
& &Y^{(0)} = (X^{(0)})^T \beta^0 + \eps^{(0)},\nonumber\\
& &\beta^0 = \argmin_{\beta} \EE|f^0(X^{(0)}) - (X^{(0)})^T \beta|^2,
\end{eqnarray}
where, due to the projection property, $\EE[\eps^{(0)} X^{(0)}] =
\Cov(\eps^{(0)},X^{(0)}) = 0$. We denote the support of $\beta_0$ by $S_0 =
\{j;\ \beta^0_j \ne 0\}$. While $\EE[\eps^{(0)}] = 0$ we 
typically have that 
$\EE[\eps^{(0)}|X^{(0)}] \neq 0$, because $\EE[\eps^{(0)}|X^{(0)}] =
f^0(X^{(0)}) - X^{(0)} \beta^0$. Thus, when conditioning on $X^{(0)}$ the
assumption of zero mean for the 
error is not valid. However, when the inference for $\beta^0$ 
is unconditional (not conditioning on $X^{(0)}$), then we have zero mean for the
error: therefore, due to model misspecification, the inference with random
design should always be \emph{unconditional} on $X^{(0)}$.

We note that $\beta^0$ still has interesting model-free (and well known)
interpretations such as: the $j$th component $\beta^0_j = L_{j} \cdot 
\mathrm{Parcorr}(Y^{(0)},X_j^{(0)}|\{X_k^{(0)};\ k \neq j\})$ equals the
partial correlation 
between $Y^{(0)}$ and $X_j^{(0)}$ given all other variables, up to a
constant $L_j = 
\sqrt{K_{jj}/K_{YY}}$, where $K^{-1}$ is the $(p+1)\times (p+1)$
covariance matrix of $(Y,X)$; thus, $\beta^0_j$ measures the linear effect
of $X_j$ on $Y$ after adjusting for the linear effects of all other
variables $X_k\ (k \neq j)$ on $Y$. In addition, for Gaussian design, we
have the following important interpretation: if $\beta^0_j \neq 0$, then
the variable $X_j^{(0)}$ is in the active set (i.e., relevant) of the
nonlinear true function $f^0$, see Proposition \ref{prop3}. 


We consider here a concrete set of assumptions for Theorem \ref{th1}
below. Denote by  
\begin{eqnarray*}
& &\gamma_j^0 = \argmin_{\gamma} \EE|X_j^{(0)} - \sum_{k\neq j} \gamma_{j,k}
X_k^{(0)}|^2,\\
& &Z_j^{(0)} = X_j^{(0)} - \sum_{k\neq j} \gamma^0_{j,k} X_k^{(0)}
\end{eqnarray*}
the population regression vector and residual variables when regressing the
random variable $X_j^{(0)}$ on all other variables $\{X_k^{(0)};\ k \neq
j\}$. It is well known that $\gamma_j^0 =
- (\Sigma^{-1})_{\bullet j}/(\Sigma^{-1})_{jj}$, where
$(\Sigma^{-1})_{\bullet j}$ denotes the $j$th column vector of
$\Sigma^{-1}$ (assuming it exists, see (A1)).  

\medskip\noindent
\textbf{Assumptions.}\\
The covariables are such that:
\begin{description}
\item[(A1)] $\Cov(X^{(0)}) = \Sigma\ \mbox{has smallest eigenvalue}\
\Lambda^2_{\mathrm{min}}(\Sigma) \ge C_1 > 0$;
\item[(A2)] $\max_{j} \|X_j^{(0)}\|_{\infty} \le C_2 < \infty$;
\item[(A3)] $\|Z_j^{(0)}\|_{\infty} \le C_3 < \infty$;
\item[(A4)] We have either:
\begin{enumerate}
\item[(a)]$\|\gamma^0_j\|_1 = o(\sqrt{n/\log(p)})$, $\|\gamma^0_j\|^r_r
 = o\left((n/\log(p)\right)^{\frac{1-r}{2}} \log(p)^{-1/2})$ for $0 < r <
  1$, and the maximal eigenvalue of $\bx_{S_j}^T \bx_{S_j}/n$ satisfies 
  $\hat{\Lambda}^2_{\mathrm{max}}(S_j) = O_P(1)$, where $\bx_{S_j}$ denotes
  the submatrix of the 
design with columns corresponding to $S_j = \{k;\ \gamma^0_{j,k} \neq 0\}$;
\end{enumerate}
or
\begin{enumerate}
\item[(b)] $s_j = |S_j| = \|\gamma^0_j\|_0^0 = \sum_{k \neq j}I((\Sigma^{-1})_{jk}
\neq 0 ) = o(\sqrt{n}/\log(p))$. 
\end{enumerate}
\end{description}

\medskip\noindent
Regarding the structure of the regression:
\begin{description}
\item[(A5)] The sparsity satisfies either:
\begin{enumerate}
\item[(a)] $\|\beta^0\|_1 = o(\sqrt{n/\log(p)})$, $\|\beta^0\|^r_r
  \hat{\Lambda}_{\mathrm{max}}^{r}(S_0) =
  o_P\left((n/\log(p)\right)^{\frac{1-r}{2}} \log(p)^{-1/2})$ for $0 < r <
  1$, and the maximal eigenvalue of $\bx_{S_0}^T \bx_{S_0}/n$ satisfies 
  $\hat{\Lambda}^2_{\mathrm{max}}(S_0) = O_P(1)$, where $\bx_{S_0}$ denotes
  the submatrix of the 
design with columns corresponding to $S_0 = \{j;\ \beta^0_j \neq 0\}$;
\end{enumerate}
or
\begin{enumerate}
\item[(b)] $s_0 = |S_0| = \|\beta^0\|_0^0 = \sum_{j=1}^p
I(\beta^0_j \neq 0) = o(\sqrt{n}/\log(p))$.
\end{enumerate}

\item[(A6)]For the second moment $\omega_{p;jj}^2 := \EE|\eps^{(0)}
  Z_{j}^0|^2$: $\omega_{p;jj}^2 \ge C_4$ for some constant $C_4 > 0$. (The
  existence of  $\omega_{p;jj} < \infty$ is implied by (A3) and (A7)). 
\item[(A7)]
 The error satisfies one of the following conditions:
\begin{enumerate}
\item[(a)]$|\eps^{(0)}| \le V$, where $V$ is a fixed random variable (not
  depending on $p$) with $\EE|V|^2 <  \infty$;
\end{enumerate}
or
\begin{enumerate}
\item[(b)] $\EE|\eps^{(0)}|^{2+\delta}  \le C_5 <  \infty$ for some $\delta
  >0$.
\end{enumerate}
Either of the conditions implies that for some constant $C_6
< \infty$, $\EE|\eps^{(0)}|^2 \le C_6 < \infty$.
%
\end{description} 
The assumptions (A2) and (A3) are somewhat restrictive (see also (B1) in
\citet{vdgetal13}). Assumption (A3) is implied by (A2) and assuming
that $\|\gamma^0_j\|_1$ is bounded. Examples where (A7) holds
are discussed in Section \ref{subsec.projection}. Regarding the assumptions
(A4) and (A5) we first note that:
\begin{eqnarray*}
& &\mbox{(A4) can be replaced by (D2) in Section \ref{subsec.prelim}},\\
& &\mbox{(A5) can be replaced by (D3) in Section \ref{subsec.prelim}},
\end{eqnarray*}
see Section \ref{subsec.proofth1} and Lemma \ref{lemm2}. Furthermore, for
$\ell_r$ sparsity in (A4,a) and (A5,a), the condition on the maximal
eigenvalue can be relaxed by requiring for 
e.g. (A5,a) that 
\begin{eqnarray*}
S^* = \{j;\ |\beta^0_j| > C
\sqrt{\log(p)/n}/\hat{\Lambda}_{\mathrm{max}}(S_0)\},
\end{eqnarray*}
for some $0 < C < \infty$, has cardinality $S^* = o(n/\log(p))$; and
analogously for condition 
(A4,a). Requiring some sparsity for the design as in (A4) is due to our
proof of Proposition \ref{prop2}: this is in contrast 
for fixed design, where no sparsity condition on the design is needed when
using the nodewise square root Lasso in (\ref{nodewisesqrtlasso}) (see
Theorem \ref{th2}). Finally, a sparsity assumption as in (A5) is typical
for the de-sparsified Lasso \citep{zhangzhang11,vdgetal13,vdg14}.

\begin{theo}\label{th1}
Consider the de-sparsified Lasso in (\ref{desplasso}) with
(\ref{nodewiselasso}) or (\ref{nodewisesqrtlasso}), and the parameter
$\beta^0$ in (\ref{betarand}) induced by the random design model
(\ref{mod.true}).    
Assume (A1)-(A7). If $\lambda = D_1 \sqrt{\log(p)/n}$ and $\lambda_X =
D_2 \sqrt{\log(p)/n}$ for $D_1, D_2$ sufficiently large, then:
\begin{eqnarray*}
\sqrt{n} \frac{Z_j^T X_j/n}{\omega_{p;jj}} (\hat{b}_j - \beta^0_j)
\Rightarrow {\cal N}(0,1)\ (n \to \infty),
\end{eqnarray*}
where $\omega_{p;jj}^2 = \EE|\eps^{(0)} Z_{j}^0|^2$ .
\end{theo}
A proof is given in Section \ref{sec.proofs}. The
representation of the normalization factor should facilitate to recognize its
order of magnitude $\sqrt{n}$. For construction of  
confidence intervals and hypothesis tests we need to consistently estimate
the quantity $\omega_{p;jj}$: this is discussed in the following Section 
\ref{subsec.estvar}. 

\bigskip\noindent
\textbf{Remark 1.} If the assumptions in (A3), (A4) and (A6) hold uniformly
in $j$, we can rephrase the statement of Theorem \ref{th1} as follows:
\begin{eqnarray*}
& &\frac{Z_j^T X_j}{\sqrt{n} \omega_{p;jj}}(\hat{b}_j - \beta^0_j) =
\Delta_j + W_j,\\
& &\max_{j=1,\ldots,p} |\Delta_j| = o_P(1),\ W_j \Rightarrow {\cal N}(0,1)
\end{eqnarray*}


\subsection{Estimation of the variance}\label{subsec.estvar}

We can estimate 
$\omega_{p;jj}^2 = \EE|\eps^{(0)} Z_{j}^0|^2$ by the empirical variance of
$\hat{\eps}_i Z_{j;i}$,
\begin{eqnarray*}
n^{-1} \sum_{i=1}^n (\hat{\eps}_i Z_{j;i} - 
  n^{-1} \sum_{r=1}^n \hat{\eps}_r Z_{j;r})^2,\ \ \hat{\eps} = Y - \bx \hat{\beta}.
\end{eqnarray*}
\begin{prop}\label{prop.variance}
Consider the random design model (\ref{mod.true}) with the projected parameter
$\beta^0$ in (\ref{betarand}). Assume (A1), (A2), (A3), $\|\beta^0\|_1 =
o(\sqrt{n/\log(p)})$ (which is part of assumption (A5)), (A6), (A7) and
(D2) from Section \ref{sec.proofs} (the latter is implied by the additional
assumption (A4)). Then,   
\begin{eqnarray*} 
\hat{\omega}_{p;jj}^2/\omega_{p;jj}^2 = 1 + o_P(1).
\end{eqnarray*}
\end{prop}
A proof is given in Section \ref{sec.proofs}. We have as an estimate of the normalizing
factor in Theorem \ref{th1} the following expression: 
\begin{eqnarray}\label{varest}
\frac{Z_j^T X_j}{\sqrt{n} \hat{\omega}_{p;jj}},
\end{eqnarray}
corresponding to the ``sandwich formula'' in the case with $p
< n$
\citep{eicker1967limit,huber1967behavior,white1980heteroskedasticity,freedman1981}. 


In particular the formula in (\ref{varest}) is different than the usual
expression for correctly specified high-dimensional linear models, used in
\citet{vdgetal13},  
\begin{eqnarray}\label{varestclassic}
\frac{Z_j^T X_j}{\|Z_j\|_2 \hat{\sigma_{\eps}}},
\end{eqnarray}
where $\hat{\sigma}_{\eps}^2$ is an estimate of the error variance
$\sigma_{\eps}^2$,  e.g., $\hat{\sigma}_{\eps}^2 = n^{-1} \sum_{i=1}^n
(\hat{\eps}_i - n^{-1} \sum_{r=1}^n \hat{\eps}_r)^2$ with $\hat{\eps} = Y -
\bx \hat{\beta}$. While the formula in
(\ref{varestclassic}) is asymptotically valid for correctly specified
models, the analogue in (\ref{varest}) is robust and valid irrespective
whether the model is correct or not. The expression in
(\ref{varest}) is now also implemented in the \texttt{R}-software package
\texttt{hdi}. 

%

\subsection{Sparsity of the projection and implications on the error $\eps^{(0)}$}\label{subsec.projection}

The statement in Theorem \ref{th1} depends, among other conditions, on
assumptions (A5)-(A7) which are depending on the projection of the
nonlinear to a linear model. In particular, (A5) requires sparsity of the
projected parameter vector: even if the underlying true nonlinear
regression function depends only on a few covariables, the projected
parameter $\beta^0$ in (\ref{betarand}) is not necessarily sparse. We
provide here some sufficient conditions ensuring a sparse $\beta^0$. 

Throughout this subsection, $\beta^0$ is as in (\ref{betarand}). We know that 
\begin{eqnarray*}
& &\beta^0 = \Sigma^{-1} \Gamma,\\
& &\Sigma = \Cov(X^{(0)}),\ \Gamma = (\Cov(f^0(X^{(0)}),X_1^{(0)}),\ldots,
\Cov(f^0(X^{(0)}),X_p^{(0)})^T. 
\end{eqnarray*}
Therefore,
\begin{eqnarray}\label{beta-form}
\beta^0_j = \sum_{\ell=1}^p (\Sigma^{-1})_{j \ell} \Gamma_{\ell}.
\end{eqnarray}
Denote by $\|\Sigma^{-1}\|_{\infty} = \max_{jk} |(\Sigma^{-1})_{jk}|$ and
by $(\Sigma^{-1})_{\bullet \ell}$ the $\ell$th column of $\Sigma^{-1}$, and
generally by $\|u\|_0^0 = \sum_{r=1}^d I(u_r \neq 0)$ the $\ell_0$-sparsity
of a $d$-dimensional vector $u$.   
\begin{prop}
\label{prop-sparsity}
Consider the random design model (\ref{mod.true}) with the projected parameter
$\beta^0$ in (\ref{betarand}). Assume that $\Sigma$ is positive definite
(but not requiring bounds on 
its eigenvalues). The following holds:
\begin{enumerate}
\item $\ell_r$-sparsity for $0 < r \le 1$: 
\begin{eqnarray*}
\|\beta^0\|_r \le \max_{\ell} \|(\Sigma^{-1})_{\bullet \ell}\|_r \|\Gamma\|_r,
\end{eqnarray*}
which implies, for $s_{\ell} = \|\gamma^0_{\ell}\|_0^0 = \sum_{k \neq \ell}
I((\Sigma^{-1})_{k \ell} \neq 0)$,
\begin{eqnarray*}
\|\beta^0\|_r \le (\max_{\ell} s_{\ell}+1)^{1/r} \|\Sigma^{-1}\|_{\infty} \|\Gamma\|_r.
\end{eqnarray*}
\item $\ell_0$-sparsity: 
\begin{eqnarray*}
\|\beta^0\|_0^0 \le \sum_{\ell \in S_{\Gamma}} (s_{\ell}+1),\ S_{\Gamma} = \{j;\ \Gamma_j \neq 0\},
\end{eqnarray*}
which implies 
\begin{eqnarray*}
\|\beta^0\|_0^0 \le (\max_{\ell} s_{\ell} +1) \|\Gamma\|_0^0.
\end{eqnarray*}
\end{enumerate}
\end{prop}
A proof is given in Section \ref{sec.proofs}. As an example, consider the
case where $\Sigma$ is block-diagonal with 
maximal block-size equal to $b_{\mathrm{max}}$. We then have that
$\max_{\ell} s_{\ell} + 1 = b_{\mathrm{max}}$ and hence by
Proposition \ref{prop-sparsity}: 
\begin{eqnarray*}
& &\|\beta^0\|_r \le b_{\mathrm{max}}^{1/r} \|\Sigma^{-1}\|_{\infty}
\|\Gamma\|_r\ \ (0 < r
\le 1),\\
& &\|\beta^0\|_0^0 \le b_{\mathrm{max}} \|\Gamma\|_0^0.
\end{eqnarray*}

\bigskip
\paragraph{Block dependence.}
Assume now that the predictor variables exhibit
block dependence with blocks corresponding to the associated block-diagonal
covariance matrix $\Sigma$. That is, there are blocks of variables, where
the variables from different blocks are
(jointly) independent, and these blocks induce a block-diagonal covariance
matrix. Denote by $S_{f^0} \subseteq \{1,\ldots ,p\}$ the
support of $f^0(\cdot)$ which contains all the variables which have an
influence in $f^0(\cdot)$. 
\begin{corr}\label{corr1}
Assume the conditions of Proposition \ref{prop-sparsity}. In addition,
assume block dependence with maximal block-size equal to
$b_{\mathrm{max}}$. We have that   
\begin{eqnarray*}
\|\Gamma\|_0^0 \le b_{\mathrm{max}}|S_{f^0}|,
\end{eqnarray*}
and, due to Proposition \ref{prop-sparsity},
\begin{eqnarray*}
\|\beta^0\|_0^0 \le b_{\mathrm{max}}^2 |S_{f^0}|.
\end{eqnarray*}
\end{corr}
A proof is given in Section \ref{sec.proofs}. 

Proposition \ref{prop-sparsity} and Corollary \ref{corr1} obviously lead to 
justifications of the assumption on the sparsity $s_0$ in (A5), but also
for the conditions in (A7). Regarding the latter: 
if $\|\beta^0\|_1 \le C_9 < \infty$ (which is implied by $\|\beta^0\|_0^0$
bounded and $\max_j |\beta^0_j|$ bounded) and assuming (A2) we
have that 
\begin{eqnarray*}
|\eps^{(0)}| = |Y^{(0)} - (X^{(0)})^T \beta^0| \le |Y^{(0)}| + C_9 C_2.
\end{eqnarray*}
Thus, assuming either $|Y^{(0)}| \le V$ for
some fixed random variable $V$ with $\EE|V|^2< \infty$ or $\EE|Y^{(0)}|^{2+\delta}
\le M_3 < \infty$ (which are both rather weak assumptions) implies either
(A7,a) or (A7,b), respectively. 
   
%

%

\subsection{Gaussian design}

The bound in Proposition \ref{prop2} and Corollary \ref{corr1} for
$\ell_0$-sparsity can be much improved when assuming that $X^{(0)}$ has a
joint Gaussian distribution. This is in conflict with assumption
(A2). However, for the case with Gaussian design, thereby dropping (A2) and
(A3), it would be easier to derive the statements from Theorem \ref{th1}
and Proposition \ref{prop.variance}.
\begin{prop}\label{prop3}
Consider the random design model (\ref{mod.true}) with the projected parameter
$\beta^0$ in (\ref{betarand}). Assume that $X^{(0)}$ has a joint Gaussian
distribution with positive definite covariance matrix $\Sigma$ (but not
requiring bounds on its eigenvalues). Then,
\begin{eqnarray*}
S_0 \subseteq S_{f^0}.
\end{eqnarray*}
\end{prop}
A proof is given in Section \ref{sec.proofs}. This is an important result
saying that if we infer a variable as an active variable (significantly
different from zero) in the misspecified linear model, it must be an active
variable in the nonlinear true model.  

To make further statements, we represent the function $f^0$ as follows:
\begin{eqnarray*}
& &f^0(x) = \sum_{k=1}^d f_k^0(x_{S_k}),\\
& &\{S_1,\ldots ,S_d\}\ \mbox{a partition:}\ S_{f^0} = \cup_{k=1}^d S_k,\
S_k \cap S_{\ell} = \emptyset\ (k \neq \ell),
\end{eqnarray*}
where $x_A$ denotes the subvector of $x$ with components in $A \subseteq
\{1,\ldots ,p\}$ and $\EE[f_k^0(X_{S_k})] = 0$; and the partition is finest
in the sense that the representation of $f^0$ 
is given with the $S_k$'s of smallest possible cardinality. For example,
for the function considered in Section \ref{sec.empirical}
\begin{eqnarray}\label{ex-f0}
f^0(x) = - 5 + 5 \sin (\pi x_1 x_2) + 4 (x_3-0.5)^2 + 2 x_5 + x_6,
\end{eqnarray}
we have the partition $S_1 = \{1,2\},\ S_2 = \{3\}, S_3 = \{5\}, S_4 =
\{6\}$. 
\begin{prop}\label{prop4}
Consider the random design model (\ref{mod.true}) with the projected parameter
$\beta^0$ in (\ref{betarand}). Assume that $X^{(0)}$ has a joint Gaussian
distribution with positive definite covariance matrix $\Sigma$ (but not
requiring bounds on its eigenvalues). Consider the projected parameter in
the submodel with 
variables from $S_k$ ($k \in \{1,\ldots ,d\}$):
\begin{eqnarray*}
\tilde{\beta}(S_k) = \argmin_{\beta \in \R^{|S_k|}} \EE|f_k^0(X^{(0)}_{S_k}) -
(X^{(0)}_{S_k})^T \beta|^2.
\end{eqnarray*}
For $j \in S_k$ we denote by $c(j)$ the index of the component in
$\tilde{\beta}(S_k)$ which corresponds to variable $X^{(0)}_j$. Then,
\begin{eqnarray*}
\beta^0_j = \tilde{\beta}_{c(j)}(S_k),
\end{eqnarray*}
saying that we can infer $\beta^0_j$ with $j \in S_k$ from the submodel
with variables $X^{(0)}_{S_k}$.
\end{prop}
A proof is given in the Appendix. As an example, we consider again $f^0$
from (\ref{ex-f0}). Proposition \ref{prop4} then implies:
\begin{eqnarray*}
& &(\beta^0_1,\beta^0_2)^T = \argmin_{\beta \in \R^2} \EE|5 \sin (\pi X^{(0)}_1 X^{(0)}_2) -
(X^{(0)}_{1},X^{(0)}_2) \beta|^2 = (0,0)^T,\\
& &\beta^0_3 = \argmin_{\beta \in \R} \EE|4(X^{(0)}_3 - 0.5)^2 - 5 - X^{(0)}_3
\beta|^2 = -4,\\
& &\beta^0_5 = \argmin_{\beta \in \R} \EE|2 X^{(0)}_5 - X^{(0)}_5
\beta|^2 = 2,\\
& &\beta^0_5 = \argmin_{\beta \in \R} \EE|X^{(0)}_5 - X^{(0)}_6
\beta|^2 = 1,
\end{eqnarray*}
and all $\beta_j^0 = 0$ for $j \notin S_{f^0}$. For the numerical values of
$\beta_1^0, \beta_2^0$ and $\beta_3^0$, we used that
$X^{(0)}$ has mean zero. 

\section{Fixed design model}\label{sec.fixeddesign}

Consider the model as in (\ref{mod.true}) but now with fixed design:
\begin{eqnarray}\label{mod.fixeddes}
Y^{(i)} = f^0(X^{(i)}) + \xi^{(i)},\ i=1,\ldots ,n,
\end{eqnarray}
where $\xi^{(1)},\ldots ,\xi^{(n)}$ are i.i.d. with $\EE[\xi^{(i)}] = 0$ and
$\EE|\xi^{(i)}|^2 = \sigma^2$. As before, we denote the $n \times p$ design
matrix by $\bx$ and the $n \times 1$ response vector by $Y = (Y^{(1)},\ldots
,Y^{(n)})^T$.  We assume that $\mathrm{rank}(\bx) = n \le p$ and thus, we
can always represent the vector 
$\mathbf{f}^0 = (f^0(X^{(1)}),\ldots ,f(X^{(n)}))^T$ as $\bx \beta^{\dagger}$. The
vector $\beta^{\dagger}$ is not unique, but we can look for some sparsest
solution. We consider the basis pursuit solution \citep{chen1998atomic},
known also as the solution from compressed sensing
\citep{candes2006near,donoho2006compressed}: 
\begin{eqnarray}\label{basis-purs}
\beta^0 = \argmin_{\beta} \{\|\beta\|_1; \bx \beta = \mathbf{f}^0\}.
\end{eqnarray}

Thus, the model in (\ref{mod.fixeddes}) is correctly specified as a linear
model 
\begin{eqnarray}\label{mod.lin}
Y = \bx \beta^0 + \eps \ \mbox{with}\ \beta^0\ \mbox{as in (\ref{basis-purs})},
\end{eqnarray}
where $\eps = (\xi_1,\ldots ,\xi_n)^T$. In particular, due to correct
specification, the interpretation of $\beta^0$ is standard.

We refer to this $\beta^0$ in (\ref{basis-purs}) throughout this
section (unless stated otherwise). We assume the following:
\begin{description}
\item[(B1)] $\lambda_X \asymp\sqrt{\log(p)/n}$ and $\|Z_j\|_2^2/n \ge C >
  0$;
\item[(B2)] $\|\hat{\beta}(\lambda) - \beta^0\|_1 = o_P(1/\sqrt{\log(p)})$.
\end{description}
We justify these assumptions below. 
\begin{theo}\label{th2}
Consider the de-sparsified Lasso in (\ref{desplasso}) with
(\ref{nodewiselasso}) or (\ref{nodewisesqrtlasso}), and the fixed design
model (\ref{mod.fixeddes}) with 
$\mathrm{rank}(\bx) = n$ and linear representation as in (\ref{mod.lin}) with
$\beta^0$ as in (\ref{basis-purs}). Assume either Gaussian 
errors or condition (A7) and assume that $\sigma^2 \ge L > 0$. Suppose that
(B1) and (B2) hold when using the nodewise Lasso (\ref{nodewiselasso}), or
only (B2) when using the 
nodewise square root Lasso (\ref{nodewisesqrtlasso}). Then
\begin{eqnarray*}
\frac{Z_j^T X_j}{\sigma \|Z_j\|_2} (\hat{b}_j -
\beta^0_j) \Rightarrow {\cal N}(0,1).
\end{eqnarray*}
\end{theo}
Proof: This follows from \citet[Th.2.1]{vdgetal13} for Gaussian errors. For
non-Gaussian errors, we invoke the Lindeberg condition and proceed as for
the proof of Theorem \ref{th1} (Proposition \ref{prop1}).\hfill$\Box$ 

\medskip
We argue first that (B1) holds with high probability. Assume the following.  
\begin{description}
\item Consider the setting where the rows of $\bx$ arise as fixed
  i.i.d. realizations of a $p$-dimensional random variable $X$ with
  covariance matrix $\Sigma$.
\item[(C1)] 
\begin{enumerate}
\item[(i)] 
$0 < C_7 \le 1/(\Sigma^{-1})_{jj} = \EE|Z_j^{(0)}|^2 \ge C_8 < \infty$ (the
upper bound is implied by (A3); the lower bound is the analogue of (A6));
\item[(ii)] $\max_j  \|X_j\|_{\infty} \le C_2 < \infty$ (which is assumption
(A2)); 
\item[(iii)] $\|\gamma^0_j\|_1  = o(\sqrt{n/\log(p)})$ (which is part of the
assumption (A4a)).
\end{enumerate}
\item[(C2)] (A1), (A2), (A5) and (A7).  
\end{description}

\begin{prop}(for nodewise Lasso only)\label{prop-Ci}
Assume that (C1) holds. Then, for $\lambda_X = D_2 \sqrt{\log(p)/n}$ with
$D_2$ sufficiently large, assumption (B1) holds with probability tending to
one.    
\end{prop}
A proof is given in Section \ref{sec.proofs}.\hfill$\Box$ 

\begin{prop}\label{prop-Cii}
Consider the fixed design model (\ref{mod.fixeddes}) having a linear
representation as in (\ref{mod.lin}) with $\beta^0$ as in
(\ref{basis-purs}). Assume that (C2) holds. Then, for 
$\lambda = D_1 \sqrt{\log(p)/n}$ with $D_1$ sufficiently large,
assumption (B2) holds with probability tending to one.     
\end{prop}
Proof. The statement can be derived as in the proof of statement 2 in Lemma
\ref{lemm2} in Section \ref{sec.proofs}.\hfill$\Box$ 
%


\paragraph{Sparse solutions and misspecification.}
We note that for a fixed design linear model, misspecification with
respect to the linearity in the unknown parameters cannot happen. The same
is true when conditioning on the covariables $X$. In this scenario, we do
not need to employ the ``sandwich'' variance formula in (\ref{varest}) but
we can use the more standard expression from (\ref{varestclassic}). 
What is
important though is the interpretation of the parameter $\beta^0$ and of
the output of the de-sparsified Lasso: the inferential statements are valid
for a sparse approximation. We focused here on the choice of the basis
pursuit solution in (\ref{basis-purs}) which is perhaps among the simplest
and which can be computed. 
But in fact, \emph{any} solution of $\bx \beta = \mathbf{f}^0$ satisfying
assumption (B2) is good enough: or in view of Proposition \ref{prop-Cii},
any solution which is weak $\ell_r$- ($0<r<1$) or $\ell_0$-sparse, see
(A5), is fine. A confidence interval then means that it
covers \emph{any} sufficiently $\ell_r$- and $\ell_0$-sparse solution
$\beta^0$ of $\bx \beta = \mathbf{f}^0$. This itself is a nice and
``strong'' interpretation of a confidence interval, namely that despite
non-uniqueness, it covers all sparse solutions.  

\section{Some empirical results}\label{sec.empirical}
We consider two non-linear models as in (\ref{mod.true}) (or versions
thereof for fixed design, see Section \ref{subsec.simul2}). The first one
uses a nonlinear regression 
function from \nocite{friedman1991} Friedman's (1991) MARS paper but with
smaller signal to noise ratio:  
\begin{description}
\item[(M1)]
\begin{eqnarray*}
& &\hspace*{-15mm}X^{(0)} \sim {\cal N}_p(0,\Sigma),\ \Sigma_{j,j} = 1\
\forall j,\ \Sigma_{3,4} = \Sigma_{4,3} = 0.8,\ \Sigma_{j,k} = 0\ (j \neq
k;\ j,k \notin \{3,4\}),\\ 
& &\hspace*{-15mm}f^0(x) = -5 + 2 \sin (\pi x_1 x_2) + 4 (x_3-0.5)^2 + 2 x_5 + x_6,\\
& &\hspace*{-15mm}\xi^{(0)} \sim {\cal N}(0,1).
\end{eqnarray*}
\item[(M2)]
\begin{eqnarray*}
& &X^{(0)}\ \mbox{as in (M1)},\\
& &f^0(x) = \sin(\pi/2 x_1) x_2 + x_3^3/5 + x_5 + x_6/2,\\
& &\xi^{(0)} \sim {\cal N}(0,1).
\end{eqnarray*}
\item[(M3)] 
\begin{eqnarray*}
X^{(0)}\ \sim\ {\cal N}_p(0,\Sigma),\ \Sigma_{j,k} = 0.8^{|j-k|},\ f^0\ \mbox{as
  in (M1)}.
\end{eqnarray*}
\item[(M4)] 
\begin{eqnarray*}
X^{(0)}\ \sim\ \mbox{as in (M3)},\ f^0\ \mbox{as in (M2)}.
\end{eqnarray*}
\end{description}
The intercept $-5$ in the function $f^0$ in (M1) and (M3) ensures that
$\EE[f^0(X^{(0)})] = 0$. 

\subsection{Simulations for random design}\label{subsec.simul1}

\begin{figure}[!htb]
\begin{center}
\includegraphics[scale=0.7]{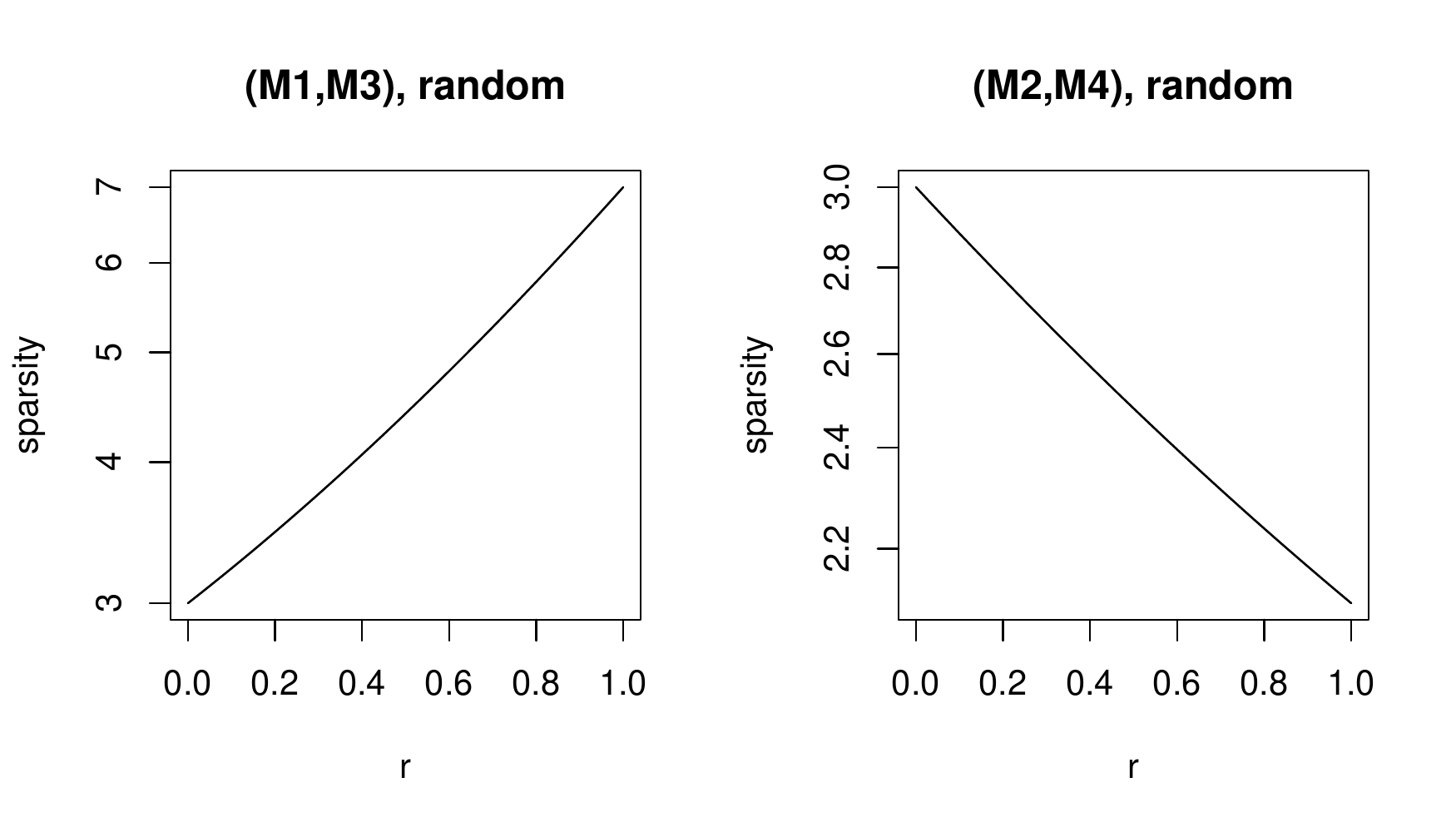}
\end{center}
\caption{Random design for models (M1),(M3) and (M2),(M4) with $p=1000$. Plot of
  $\|\beta^0\|_r^r$ (on log-scale) as a function of $r \in [0,1]$ ($r = 0$
  corresponds to the $\ell_0$-sparsity), where $\beta^0$ is as in
  (\ref{betarand}).}\label{fig.randspars} 
\end{figure}

For random design, the corresponding parameters $\beta^0$ in
(\ref{betarand}) are as follows: 
\begin{eqnarray*}
& &\mbox{for model (M1),(M3):}\ \ \beta^0 = (0,0,-4,0,2,1,0,\ldots ,0)^T\\
& &\mbox{for model (M2),(M4):}\ \ \beta^0 = (0,0,0.6,0,1, 0.5, 0, \ldots, 0)^T.
\end{eqnarray*}
The values are in accordance with Proposition \ref{prop4}, because of
Gaussianity of the design: the active set $S_0 = \{3,5,6\} \subset S_{f^0}
= \{1,2,3,5,6\}$. 
Figure \ref{fig.randspars} displays $\|\beta\|_r^r$ as a
function of $r$ for $0 \le r \le 1$.
The log-sparsity
is approximately a linear function in $r$, once increasing (for (M1),(M3))
and once decaying (for (M2),(M4)). 
Our theory requires either weak $\ell_r$-sparsity or $\ell_0$-sparsity of
$\beta^0$ (see (A5,a) or (A5,b)) and hence a possibly more realistic
assumption than $\ell_0$-sparsity alone. 

For simulations with random design, we generate $n$ independent data points
according to the models (M1)-(M4) where for each realization, we generate the
$X$ and $\xi$ variables anew. We consider the case with sample size $n=200$
and dimension $p=1000$. We 
use the de-sparsified Lasso procedure as described in (\ref{desplasso})
with the nodewise Lasso (\ref{nodewiselasso}) and tuning parameters
$\lambda$ and $\lambda_X$ (the same for all $j$) from the default in the
\textrm{R}-software package \texttt{hdi} \citep{hdipackage}. For
estimation of the asymptotic variance we use (\ref{varest}). 

\begin{figure}[!htb]
\begin{center}
  \includegraphics[scale=0.7]{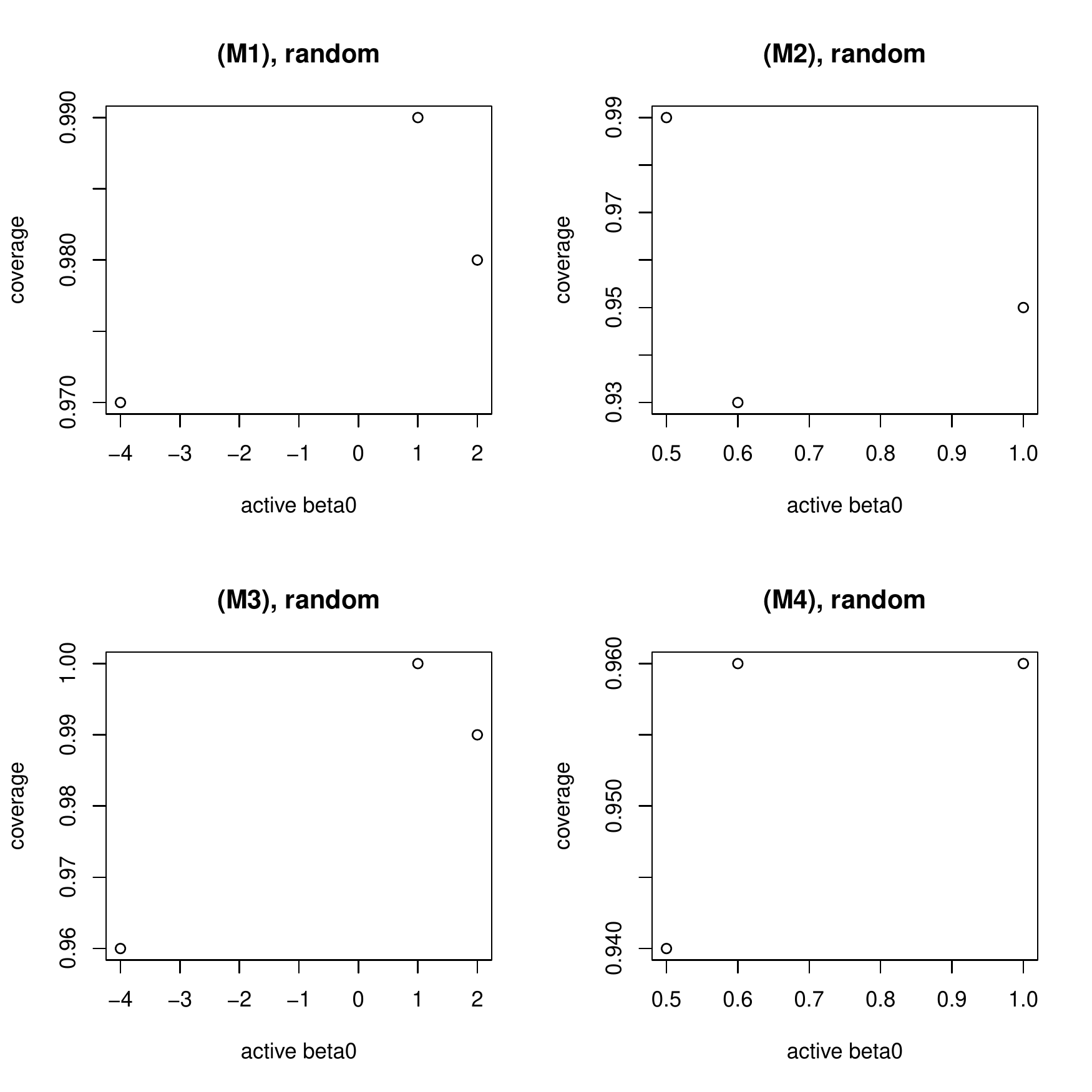}
\end{center}
\caption{Random design. Coverage as a function of the coefficients
  $\beta_j^0$ of the active variables with $j \in S_0$. Nominal level equal to
  0.95. Sample size 
  $n=200$ and dimension $p=1000$.}\label{fig1}
\end{figure}

Table \ref{tab1} and Figure \ref{fig1} report empirical results based on
100 independent  
simulations. Denoting by $\mathrm{CI}_j$ a confidence interval for
$\beta^0_j$, the average coverage is 
\begin{eqnarray}\label{avgcov}
& &\mathrm{avgcov}(S_0) = |S_0|^{-1} \sum_{j \in S_0} \PP[\beta^0_j \in
\mathrm{CI}_j],\nonumber\\
& &\mathrm{avgcov}(S_0^c) = |S_0^c|^{-1} \sum_{j \in S_0^c} \PP[\beta^0_j \in
\mathrm{CI}_j],
\end{eqnarray}
and the empirical analogue by replacing the probability ``$\PP$'' by an
empirical average over the 100 simulations. We consider the average
expected length of the confidence intervals
\begin{eqnarray}\label{avglen}
& &\mathrm{avglen}(S_0) = |S_0|^{-1} \sum_{j \in S_0}
\EE[\mbox{length}(\mathrm{CI}_j)],\nonumber\\ 
& &\mathrm{avglen}(S_0^c) = |S_0^c|^{-1} \sum_{j \in S_0^c}
\EE[\mbox{length}(\mathrm{CI}_j)], 
\end{eqnarray}
and the empirical analogue by replacing the expectation ``$\EE$'' with an
empirical average. 
\begin{table}[!htb]
\begin{center}
\begin{tabular}{c||c|c|c|c}
model & avg. coverage $S_0$ & avg. coverage $S_0^c$ & avg. length $S_0$ & avg. length $S_0^c$ \\
\hline
(M1) & 0.98 & 0.99 & 3.01 & 2.19 \\
(M2) & 0.91 & 0.95 & 0.48 & 0.41 \\
(M3) & 0.98 & 0.99 & 4.18 & 3.56 \\
(M4) & 0.95 & 0.95 & 0.70 & 0.65 
\end{tabular}
\end{center}
\caption{Random design. Average coverage and average length of confidence
  intervals 
  (empirical versions of (\ref{avgcov}) and (\ref{avglen})), for
  $S_0$ and $S_0^c$ separately (note that $S_0^c =
  \emptyset$ for (M3) and (M4)). Nominal level equal to 0.95. Sample size
  $n=200$ and dimension $p=1000$.}\label{tab1} 
\end{table}
The actual coverage results in Table \ref{tab1} and the more detailed view
given in Figure \ref{fig1} are very satisfactory.
We note that the lengths of the confidence intervals are not constant for
the same covariance model for $X$. The reason is that at least
asymptotically (see Theorem \ref{th1}), the length depends, among other
things, on $\EE|Z_j^{(0)} \eps^{(0)}|^2$, and the error term
$\eps^{(0)}$ itself depends on the true function $f^0$. This is in contrast to
fixed design, where the asymptotic length of the confidence intervals is a
function of $\EE|Z_j^{(0)}|^2 = 1/(\Sigma^{-1})_{jj}$ and $\sigma^2 = \EE|\xi_i|^2 =
\EE|\eps_i|^2$ only (see Theorem
\ref{th2} and formula (\ref{add1a}) and (\ref{add3})).

%

\subsection{Simulations for fixed design}\label{subsec.simul2}

We consider the same models (M1)-(M4) but now with fixed design with $n =
200$ and $p=1000$, where we use a
fixed realization of the $X$ variables in the corresponding model. We
generate $n$ independent data points according to the models (M1)-(M4)
where for each realization, we generate only the $\xi$ error variables anew.

\begin{figure}[!htb]
\begin{center}
\includegraphics[scale=0.7]{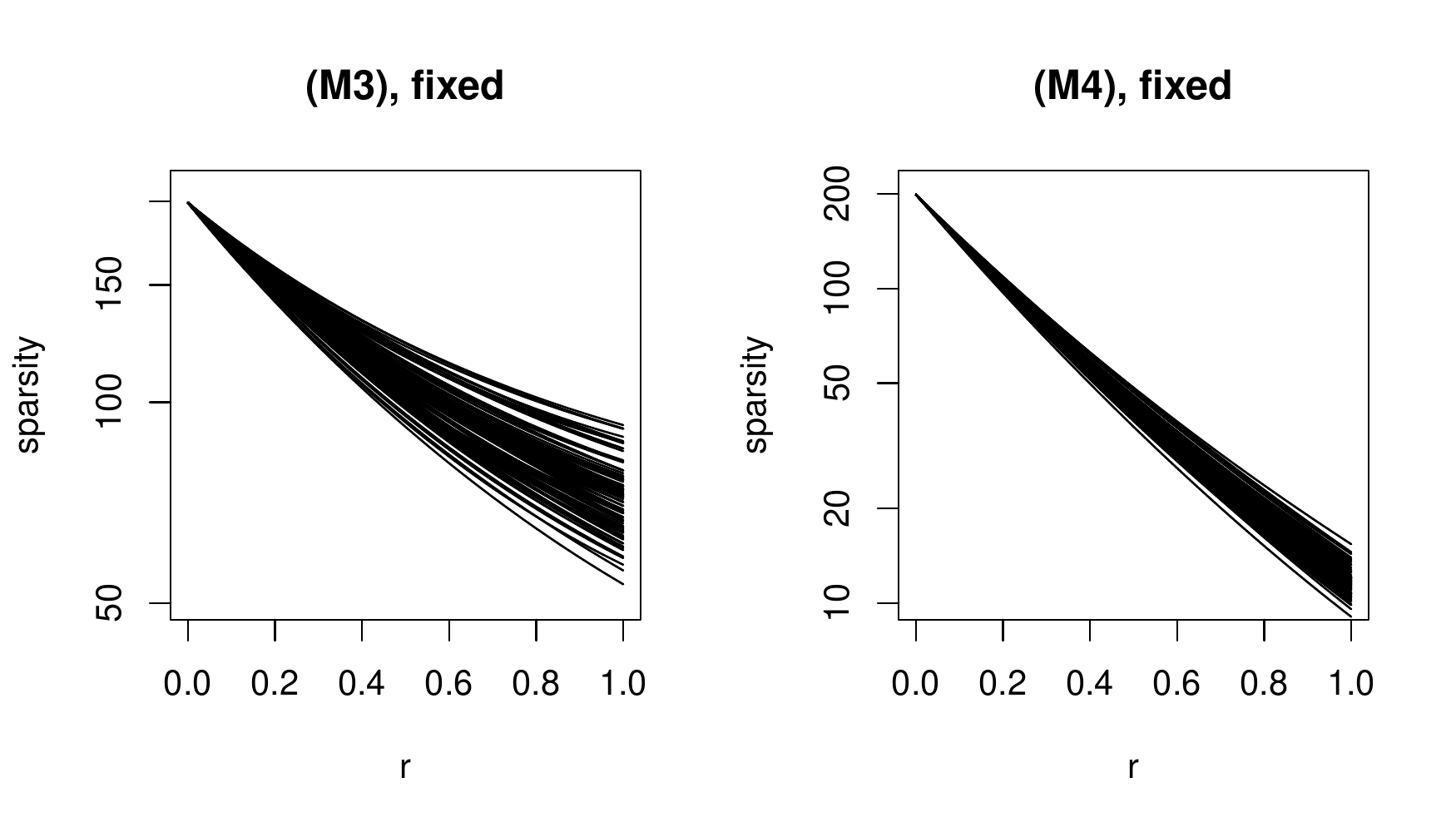}
\end{center}
\caption{Fixed design for models (M3) and (M4) with $n=200$ and
  $p=1000$. 100 independent 
  realizations and corresponding basis pursuit solutions $\beta^0$ as in
  (\ref{basis-purs}): the lines correspond to the 100 different values
  of $\|\beta^0\|_r^r$ (on log-scale) as a
  function of $r \in [0,1]$ ($r = 0$ corresponds to the
  $\ell_0$-sparsity).}\label{fig.fixedspars} 
\end{figure}

\begin{figure}[!htb]
\begin{center}
  \includegraphics[scale=0.7]{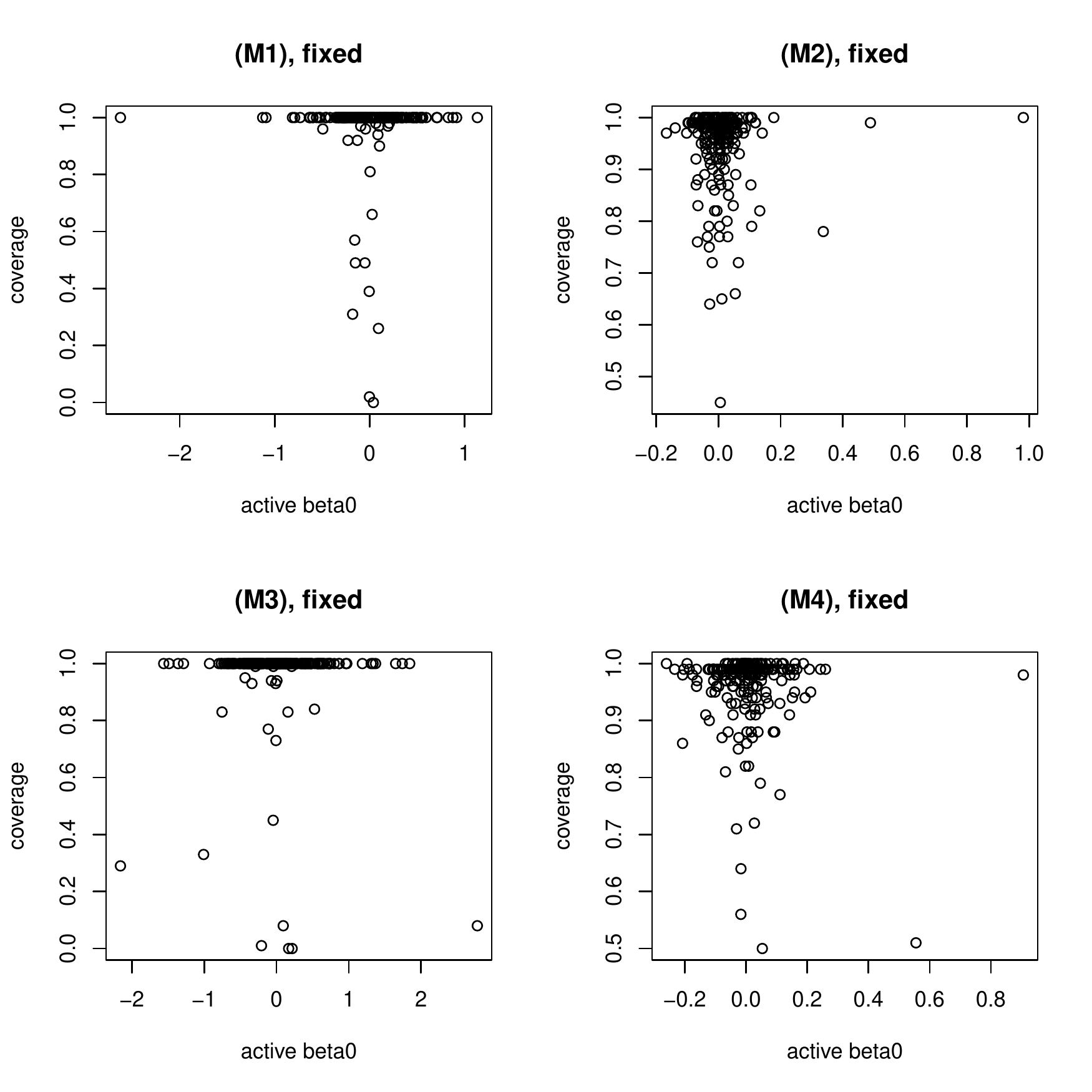}
\end{center}
\caption{Fixed design. Coverage as a function of the coefficients
  $\beta_j^0$ (from basis pursuit in (\ref{basis-purs})) of the active
  variables with $j \in 
  S_0$. Nominal level equal to 0.95. Sample size 
  $n=200$ and dimension $p=1000$.}\label{fig2}
\end{figure}

We note that for all the four models with fixed
design we have that $|S_0| = n = 200$. Figure \ref{fig.fixedspars}
displays $\|\beta^0\|_r^r$ as a function of $r$ for $0 \le r \le 1$,
where $\beta^0$ is the basis pursuit solution from (\ref{basis-purs}) and
the parameter of interest, for 100
different independent simulation runs. The log-sparsity
is approximately a linear decreasing function in $r$.
Even more pronounced here for fixed than random design, we conclude that weak
$\ell_r$-sparsity, as required by our theory, seems to be a much more realistic
assumption than $\ell_0$-sparsity which is always equal to $n =
200$. However, we also see that for model (M3), the parameter
  $\beta^0$ is not very $\ell_r$-sparse. Thus, it might be difficult
  that a confidence interval would achieve good coverage, see also Figure
  \ref{fig2} and the last paragraph of this section. 

We use the de-sparsified Lasso
procedure as described in (\ref{desplasso}) with the nodewise Lasso
(\ref{nodewiselasso}) and tuning parameters $\lambda$ and $\lambda_X$ (the
same for all $j$) from the default in the \textrm{R}-software 
package \texttt{hdi} \citep{hdipackage}. For
estimation of the asymptotic variance we use (\ref{varestclassic}). 
Table \ref{tab2} and Figure \ref{fig2} report empirical results for the
basis pursuit solution $\beta^0$ in (\ref{basis-purs}), based on 100 
independent simulations where the design is a fixed realization from the
models (M1)-(M4). 
\begin{table}[!htb]
\begin{center}
\begin{tabular}{c||c|c|c|c}
model & avg. coverage $S_0$ & avg. coverage $S_0^c$ & avg. length $S_0$ & avg. length $S_0^c$ \\
\hline
(M1) & 0.97 & 0.98 & 1.68 & 1.69 \\
(M2) & 0.95 & 0.97 & 0.41 & 0.41 \\
(M3) & 0.96 & 0.97 & 3.26 & 3.27 \\
(M4) & 0.96 & 0.96 & 0.95 & 0.95
\end{tabular}
\end{center}
\caption{Fixed design. Average coverage and average length of confidence
  intervals (empirical versions of (\ref{avgcov}) and (\ref{avglen})) for
  the basis pursuit solution $\beta^0$ in (\ref{basis-purs}), for
  $S_0$ and $S_0^c$ separately. Nominal level equal to 0.95. Sample size
  $n=200$ and dimension $p=1000$.}\label{tab2} 
\end{table}
The actual average coverage results in Table \ref{tab2} are very
fine. However, with the more detailed view in Figure \ref{fig2}, 
the coverage can be quite poor for a few coefficients although this
  should be interpreted cautiously, as explained below. The poor coverage
is particularly 
visible for the models (M1) and (M3): a reason might be that the degree of
weak $\ell_r$-sparsity of the basis pursuit solution $\beta^0$ in
(\ref{basis-purs}) is not as high as for (M2) and (M4) ((shown for
(M3), (M4) in Figure \ref{fig.fixedspars}). Regarding the
lengths of the confidence intervals: we cannot confirm the asymptotic
behavior saying that they are equal for the same covariance model for the
realized $\bx$ and the same error variances (e.g. (M1) and (M2)),
regardless of the true underlying nonlinear regression function. 

It is important to interpret the obtained confidence intervals as
  described in the last paragraph of Section \ref{sec.fixeddesign}: any
  solution of $\bx \beta 
=  \mathbf{f}^0$ which is weak $\ell_r$-sparse $(0 < r < 1)$ or
$\ell_0$-sparse is fine and should be covered by the confidence
interval. Our findings in Figure \ref{fig2} are for the basis 
pursuit solution only, and the latter is not very sparse (see Figure
\ref{fig.fixedspars}). This doesn't imply though that 
there isn't another solution $\beta^0$ which is $\ell_r$- or
$\ell_0$-sparse and whose components would be covered well by the obtained
confidence intervals. Unfortunately, the latter statement is uncheckable due to
the involved computational complexity; in contrast to the findings for the
basis pursuit solution which can be easily computed with a linear
program. Therefore, the somewhat negative findings indicated in Figure
\ref{fig2} should be down-weighted. 

\section{Discussion}\label{sec.disc}

The current work offers a precise description of interpretation and
(sufficient) assumptions for inference in a misspecified high-dimensional
linear model. The following Table \ref{tab3} summarizes the main points
with respect to interpretation and modification of the de-sparsified Lasso
procedure. A modification of the variance as in (\ref{varest}) is needed
for the case of a random design misspecified model. Such a
modification seems always advisable for the random design case, as it is
consistent irrespective whether the model is correct or not and hence
offers some robustness against model misspecification; see for example
\citet{huber1967behavior}.  
The conceptual parts, as indicated in Table \ref{tab3}, will not change for
generalized linear models as one can link them to weighted linear
regression. One should decide beforehand, whether the inference should be
performed with fixed $\bx$ (or conditional on $\bx$) or whether $\bx$ is
considered as 
random. The interpretation of the parameter $\beta^0$ (see Table
\ref{tab3}) changes when the true underlying regression function is
non-linear, perhaps more dramatically than expected. For the special case
of Gaussian random design we have the interesting property that $S_0
\subseteq S_{f^0}$ (Proposition \ref{prop3}), saying that if a variable is
significant in the misspecified linear model, it must be relevant in the
true nonlinear model.  

\begin{table}[!htb]
\begin{center}
\begin{tabular}{l|l|l}
design & interpretation of $\beta^0$ & modification \\
\hline \hline
random design & via projection in (\ref{betarand}); &
modified variance in (\ref{varest}) \\
& with model-free interp. described after (\ref{betarand});  & \\
& for Gaussian des.: active set property (Prop. \ref{prop3}) & \\
\hline
fixed design &  any sparse solution of $\bx \beta =  \mathbf{f}^0$ & no
modification\\
 &(e.g. basis pursuit solution in (\ref{basis-purs}));& \\
& with standard interp. (since no misspecif.)& \end{tabular}
\end{center}
\caption{Conceptual summary of interpretation and required modification of
  the de-sparsified Lasso procedure for misspecified high-dimensional
  linear model. The required assumptions for
  asymptotic validity of the method are described in Theorems \ref{th1} and
\ref{th2}. In case of fixed design where the true underlying regression
function is 
linear with a corresponding sparsest ``true''  parameter vector $\beta^0$,
the basis pursuit solution typically coincides with $\beta^0$ (see
compressed sensing literature \citep[cf.]{cantao07}).}\label{tab3} 
\end{table}

\subsection{Sample splitting methods}\label{subsec.samplesplit}
Regarding other methods for construction of p-values and confidence
intervals, we briefly discuss sample splitting 
techniques. Such procedures, including the preferred multiple sample instead of
single sample splitting \citep{memepb09}, can be used for the random design
misspecified case. The reason is that the sample splitting device
implicitly assumes the same probability distribution in split samples, and
this holds for random $\bx$ (but typically not for fixed $\bx$) and implies
the same projected parameter $\beta^0$ in (\ref{betarand}) in split
samples. If the linear model is correct with the same sparse true
$\beta^0$ for every sample point, sample 
splitting can also be used for fixed design cases (because both split
samples are from a fixed design linear model with parameter vector
$\beta^0$). However, for the fixed design model as in (\ref{mod.lin}), the
issue is different since e.g. the basis pursuit solution $\beta^0$ in 
(\ref{basis-purs}) would be different for every split sample.

A modification is necessary though for the misspecified random design case: even
for low-dimensional inference, which is what is used after screening for
variables in the first half of the sample, one has to use a modified
estimator for the variance, analogously to the estimator in
(\ref{varest}) which is robust against model misspecification.

\section{Proofs}\label{sec.proofs}

\subsection{Proof of Theorem \ref{th1} for random
  design}\label{subsec.proofth1} 

We prove here the statement of Theorem \ref{th1} under slightly weaker
assumptions than in condition (A). In this section, $\bx$ is always random
and the parameter $\beta^0$ as in (\ref{betarand}). 

\subsubsection{Preliminary results}\label{subsec.prelim}
We show here that the following conditions hold:
\begin{description}
\item[(D1)] $\max_{k \neq j}|\eps^T \bx_k/n| = O_P(\sqrt{\log(p)/n})$.
\item[(D2)] For either the nodewise Lasso in
  (\ref{nodewiselasso}) or the square root Lasso in
  (\ref{nodewisesqrtlasso}): $\|\hat{\gamma}_j(\lambda_X) - \gamma_j^0\|_1 =
  o_P(1/\sqrt{\log(p)}))$. 
\item[(D3)] $\|\hat{\beta}(\lambda) - \beta^0\|_1 =
  o_P(1/\sqrt{\log(p)})$.
\end{description}

\begin{lemm}\label{lemm1}
For random $\bx$, assume (A2) and $\EE|\eps^{(0)}|^2 \le C < \infty$ for
some constant $C>0$ (the latter is implied by (A7)). Then, (D1) holds,
that is: 
\begin{eqnarray*}
\max_{k \neq j}|\eps^T \bx_k/n| = O_P(\sqrt{\log(p)/n}). 
\end{eqnarray*}
\end{lemm}
Proof: Using Nemirovski's inequality \citep[Lemma 14.24]{pbvdg11} we obtain:
\begin{eqnarray*}
\EE[\max_{1 \le j \le p} |n^{-1} \eps^T X_j|^2] \le 8 \log(2p)
C_2^2 C_6 /n = O(\log(p)/n).
\end{eqnarray*}
Thus, since $\EE[\eps^T X_j] = 0$ and using Markov's inequality: 
\begin{eqnarray*}
\PP[\max_{j=1,\ldots ,p} |n^{-1} \eps^T X_j| > c] \le \EE[\max_{j=1,\ldots
  ,p} |n^{-1} \eps^T X_j|]/c \le \sqrt{\EE[\max_{j=1,\ldots
  ,p} |n^{-1} \eps^T X_j|^2]}/c = O(\sqrt{\log(p)/n})/c.
\end{eqnarray*}
This completes the proof.\hfill$\Box$

\begin{lemm}\label{lemm2}
For random $\bx$, assume (A1) and (A2). 
\begin{enumerate}
\item Then, for $\lambda_X = D_2 \sqrt{\log(p)/n}$ with
$D_2$ sufficiently large, (A3) and (A4) imply (D2).
\item If $\EE|\eps^{(0)}|^2 \le C < \infty$ for some constant $C>0$ (the
  latter is implied by (A7)), then for $\lambda = D_1 \sqrt{\log(p)/n}$ with
$D_1$ sufficiently large, (A5) implies (D3). 
\end{enumerate}
\end{lemm}
Proof: The first and second statement can be proved analogously. For the
first one, due to (A3), the error when regressing $X_j$ versus $X_{-j} =
\{X_k;\ k \neq j\}$ is bounded.

When invoking the $\ell_0$-sparsity assumptions (A4,b) or
(A5,b), respectively, we know 
that the compatibility condition holds with probability tending to one:
because of (A1), (A2) and the $\ell_0$-sparsity assumption
\citep[cf. Ch. 6.12]{pbvdg11}). Therefore, and using Lemma \ref{lemm1},
we obtain the statements invoking some oracle inequality for the Lasso
\citep[cf. Th.6.1]{pbvdg11} or the square root Lasso
\citep[Th.1.4.2]{vdg14}. 

When invoking the $\ell_r$-sparsity ($0<r<1$) assumptions (A4,a) or
(A5,a), respectively, we can use the results from \citet[Sec.5]{vdg15}
which apply not only for the square root Lasso but also for the Lasso
\citep[cf.Th.1.3.2]{vdg14}. We need to argue that the compatibility
condition holds with probability tending to one for, e.g. when proving the
second statement, the set: 
\begin{eqnarray*}
S^* = \{j;\ |\beta^0_j| > C \sqrt{\log(p)/n}/\hat{\Lambda}_{\mathrm{max}}(S_0)\}
\end{eqnarray*}
Due to the assumption on $\ell_1$-sparsity and due to the assumption that
$\hat{\Lambda}(S_0)$ is bounded, we have that $|S^*| =
o(n/\log(p))$. Therefore, due to (A1) and (A2), the compatibility condition
holds for $S^*$ with probability tending to one
\citep[cf. Ch. 6.12]{pbvdg11}).\hfill$\Box$ 
%

\subsubsection{Proof}

Denote by $Z_j^0 = X_j - \bx_{-j} \gamma_j^0$, analogously as in Section
\ref{sec.randomdes} but now for $n \times 1$ vectors. 
We first analyze the behavior of the part $Z_j^T \eps/n$. 
We have that 
\begin{eqnarray*}
\EE[\eps_i X_{k;i}] =  0\ \forall k,
\end{eqnarray*}
and hence $\EE[(Z_j^0)^T \eps] = 0$.

\begin{prop}\label{prop1}
Assume (A1), (A3), (A6) (only that $\omega_{p;jj} > 0$) and (A7). Denote
by $\omega_{p;jj}^2 = \EE|\eps^{(0)} Z_{j}^{(0)}|^2$. Then:  
\begin{eqnarray*}
\sqrt{n}\frac{\eps^T Z_j^0/n}{\omega_{p;jj}} \Rightarrow {\cal N}(0,1)\ (n
\to \infty).
\end{eqnarray*}
Note that $p = p_n$  is allowed to depend on $n$. 
\end{prop}

Proof. Denote by $W_{p;i} = \eps_i
Z_{j;i}^0$. Since $\Cov(\eps_i,X_{k;i}) = 0\ \forall k$, we have that
$\EE[W_{p;i}] = 0$. Furthermore, $W_{p;1},\ldots ,W_{p;n}$ are independent.  
We verify the Lindeberg condition. For $\kappa > 0$,
\begin{eqnarray*}
\lim_{n \to \infty}  \frac{1}{\omega_{p;jj}^2} \int_{|W_{p}| > \kappa
  \sqrt{n} \omega_{p;jj}} W_p^2 dP = 0.
\end{eqnarray*}
Assuming (A7,a), we invoke the dominated convergence theorem: 
\begin{eqnarray*}
|W_{p}|^2 I_{|W_{p}| > \kappa \sqrt{n} \omega_{p;jj}}  \le|W_{p}|^2 \le
|\eps^{(0)}|^2 |Z_j^{(0)}|^2 \le V^2 C_3^2.
\end{eqnarray*}
Because $I(|W_{p}| > \kappa \sqrt{n} \omega_{p;jj}) = 0\ (n \to \infty)$ in
probability, 
and hence 
\begin{eqnarray*}
|W_{p}|^2 I_{|W_{p}| > \kappa \sqrt{n} \omega_{p;jj}} = o_P(1),
\end{eqnarray*} 
and because of the dominated convergence theorem we conclude that the
Lindeberg condition holds. 
Assuming (A7,b), we have that $\EE|W_{p;i}|^{2 + \delta} \le
\EE|\eps_i|^{2+\delta} C_3^{2+\delta} \le  C_5
C_3^{2+\delta}$. The Lindeberg condition is then implied by the
Lyapunov theorem.\hfill$\Box$

\begin{prop}\label{prop2} (with $Z_j$ instead of $Z_j^0$)\\
Assume (A1), (A3), (A6), (A7), (D1) and
(D2). Then:   
\begin{eqnarray*}
\sqrt{n}\frac{\eps^T Z_j/n}{\omega_{p;jj}} \Rightarrow {\cal N}(0,1)\ (n
\to \infty).
\end{eqnarray*}
\end{prop}
Proof. We only need to control the difference $\eps^T (Z_j - Z_j^0)/n$. 
We have that 
\begin{eqnarray*}
|\eps^T(Z_j^0 - Z_j)/n| \le  \max_{k \neq j}|\eps^T \bx_k/n|\ \|\hat{\gamma}_j -
\gamma_j^0\|_1.
\end{eqnarray*}
The statement then follows from Proposition \ref{prop1} and invoking (D1)
and (D2).\hfill$\Box$  

%
%


\begin{prop}\label{prop-th1}
Assume (A2), (A3), (A6), (A7), (D1), (D2) and (D3). Then: 
\begin{eqnarray*}
\sqrt{n}\frac{Z_j^T X_j/n}{\omega_{p;jj}} (\hat{b}_j - \beta^0_j)
\Rightarrow {\cal N}(0,1)\ (n \to \infty).
\end{eqnarray*}
\end{prop}
Proof. The statement follows by standard arguments as in \citet{vdgetal13},
requiring (D3), and using Proposition \ref{prop2}. For the case with
  the square root Lasso in (\ref{nodewisesqrtlasso}), the proof is
  analogous. One can easily show that
  $\|Z_j\|_2/\sqrt{n} = \sqrt{\EE|Z_j^{(0)}|^2} + o_P(1)$, due to
  (A2), (A3), and (D2), and $\EE|Z_j^{(0)}|^2$ is upper bounded by
  (A3).\hfill$\Box$
 
\medskip
Using the results from Section \ref{subsec.prelim} and Proposition
\ref{prop-th1} establish the result from Theorem \ref{th1}.\hfill$\Box$

\subsection{Proof of Proposition \ref{prop.variance}}

We write 
\begin{eqnarray*}
n^{-1} \sum_{i=1}^n (\hat{\eps}_i Z_{j;i})^2 = n^{-1} \sum_{i=1}^n (\eps_i
+ (\hat{\eps}_i - \eps_i))^2 (Z_{j;i}^0 + (Z_{j;i} - Z_{j;i}^0))^2.
\end{eqnarray*}
We then get
\begin{eqnarray*}
n^{-1} \sum_{i=1}^n (\hat{\eps}_i Z_{j;i})^2 = n^{-1} \sum_{i=1}^n (\eps_i
Z_{j;i}^0)^2 + \Delta.
\end{eqnarray*}
One can easily show that $\Delta = o_P(1)$ by using H\"older's inequality
(for $\ell_1-\ell_{\infty}$; and Cauchy-Schwarz for $\ell_2-\ell_2$) 
and invoking the following:
\begin{eqnarray*}
& &\max_i|Z_{j;i}^0| \le C_3 < \infty\ \mbox{due to (A3)}\\
& &\max_i |Z_{j;i} - Z_{j;i}^0 | \le \max_i|\bx_{i,j}| \|\hat{\gamma}_j -
\gamma^0_j\|_1 = o_P(1)\ \ \mbox{due to (A2) and (D2)},\\
& &\|\hat{\eps} - \eps\|_2^2/n = \|\bx(\hat{\beta} - \beta^0)\|_2^2/n =
o_P(1)\ \mbox{due to (A2), $\|\beta^0\|_1 = o(\sqrt{n/\log(p)})$ and (A7)},
\end{eqnarray*}
where the last bound follows from e.g. \citet[Cor.6.1]{pbvdg11}. Therefore,
\begin{eqnarray*}
n^{-1} \sum_{i=1}^n (\hat{\eps}_i Z_{j;i})^2 = \EE|\eps^{(0)} Z_j^{(0)}|^2 + o_P(1).
\end{eqnarray*}
Furthermore, and simpler to obtain: 
\begin{eqnarray*}
n^{-1} \sum_{i=1}^n \hat{\eps}_i Z_{j;i}  = \EE[\eps^{(0)} Z_j^{(0)}] + o_P(1) = o_P(1).
\end{eqnarray*}
Due to (A6), the latter two displayed formulae complete the proof.\hfill$\Box$
 
\subsection{Proof of Proposition \ref{prop-sparsity}}
 
For statement 1, consider:
\begin{eqnarray*}
& &\sum_{j=1}^p |\beta^0_j|^r\\
&\le& \sum_{j=1}^p (\sum_{\ell=1}^p
|(\Sigma^{-1})_{j\ell}| |\Gamma_{\ell}| )^r \le \sum_{j=1}^p \sum_{\ell=1}^p
|(\Sigma^{-1})_{j\ell}|^r |\Gamma_{\ell}|^r = \sum_{\ell=1}^p
\|(\Sigma^{-1})_{\bullet \ell}\|_r^r |\Gamma_{\ell}|^r \le \max_{\ell}
\|(\Sigma^{-1})_{\bullet \ell}\|_r^r \|\Gamma\|_r^r. 
\end{eqnarray*}
Furthermore, we have that $\max_{\ell} \|(\Sigma^{-1})_{\bullet \ell}\|_r^r \le
(\max_{\ell} s_{\ell} + 1) \|\Sigma^{-1}\|_{\infty}^r$ and therefore
statement 1 is complete. 

Regarding statement 2, we use the following argument. Every 
point $\ell \in S_{\Gamma}$ can lead to at most $s_{\ell} + 1$
non-zero values of the components of $\beta^0$, due to formula
(\ref{beta-form}). Hence  we obtain both bounds for
$\|\beta^0\|_0^0$.\hfill$\Box$ 

\subsection{Proof of Corollary \ref{corr1}}

The bound above for $\|\Gamma\|_0^0$ follows by a similar argument as for
statement 2. in Proposition \ref{prop-sparsity}: every support point in
$S_{f^0}$ exhibits a dependence with at most $b_{\mathrm{max}}$
$X$-variables: therefore there are at most $b_{\mathrm{max}} |S_{f^0}|$
non-zero covariances between $f^0(X)$ and the $X$-variables.\hfill$\Box$

\subsection{Proof of Proposition \ref{prop3}}

It is well known that 
\begin{eqnarray*}
\beta_j^0 = \EE[Z^0_j f^0(X)] = \EE[Z^0_j f(X_{S_{f^0}})].
\end{eqnarray*}
Furthermore, since $Z_j^{(0)}$ is the residual when projecting $X^{(0)}_j$ onto
$X^{(0)}_{-j} = \{X^{(0)}_k;\ k \neq j\}$ and due to the Gaussian
assumption: $Z_j^{(0)}$ is independent of $\{X^{(0)}_k;\ k \neq j\}$. 

Therefore, if $j \notin S_{f^0}$, $Z_j^{(0)}$ is independent also of
$X^{(0)}_{S_{f^0}}$ and therefore, using the representation for $\beta^0_j$
above: $\beta^0_j = \EE[Z_j^{(0)}] \EE[f^0(X_{S_{f^0}})] = 0$, saying that $j
\notin S^0$. This proves the 
claim.\hfill$\Box$ 

\subsection{Proof of Proposition \ref{prop4}}

As mentioned already in the proof of Proposition \ref{prop3} we know that
$Z_j^{(0)}$ is independent of $\{X^{(0)}_k;\ k \neq j\}$. Therefore, for $j \in
S_k$:
\begin{eqnarray*}
\beta^0_j = \EE[Z_j^{(0)} f^0(X^{(0)})] = \EE[Z_j^{(0)} \left(f^0_1(X_{S_1}^{(0)}) +
\ldots + f^0_d(X_{S_d}^{(0)})\right)] = \EE[Z_j^{(0)} f^0_k(X_{S_k}^{(0)})].
\end{eqnarray*}
This means that we can obtain $\beta^0_j$ from projecting
$f^0(X_{S_k}^{(0)})$ onto $\{X_j^{(0)};\ j=1,\ldots ,p\}$:
\begin{eqnarray}\label{proj-all}
\gamma = \argmin_{\beta \in \R^p} \EE|f^0(X_{S_k}^{(0)}) - (X^{(0)})^T \beta|^2,
\end{eqnarray}
and $\beta^0_j = \gamma_j$. 
But we know from Proposition \ref{prop3} that for the support of $\gamma$:
\begin{eqnarray*}
S(\gamma) = \{j;\ \gamma_j \neq 0\} \subseteq S_k.
\end{eqnarray*}
Therefore, we can restrict the projection in (\ref{proj-all}) to the
variables from $S_k$:
\begin{eqnarray*}
\tilde{\gamma} = \argmin_{\beta \in \R^{|S_k|}} \EE|f^0(X_{S_k}^{(0)}) -
(X_{S_k}^{(0)})^T \beta|^2, 
\end{eqnarray*}
and $\beta^0_j = \tilde{\gamma}_{c(j)}$, where $c(j)$ the index of the
component in $\tilde{\gamma}$ which corresponds to variable
$X^{(0)}_j$. This completes the proof.\hfill$\Box$ 

\subsection{Proof of Proposition \ref{prop-Ci}}


We write
\begin{eqnarray}\label{add1a}
& &\|Z_j\|_2^2/n = \|Z_j^0\|_2^2/n + \|\bx_{-j}(\hat{\gamma}_j -
\gamma_j^0) \|_2^2/n + \Xi,\nonumber\\
& &\ |\Xi| \le 2 \|Z_j^0\|_2/\sqrt{n} \|\bx_{-j}(\hat{\gamma}_j -
\gamma_j^0) \|_2/\sqrt{n}. 
\end{eqnarray}
Due to (C1,i) we have that 
\begin{eqnarray}\label{add2}
\|Z_j^0\|_2^2/n \ge C_7/2\ \mbox{with probability tending to one}.
\end{eqnarray}
We can also establish, analogous to \citet[Cor.6.1]{pbvdg11} invoking
(C1,iii), but now controlling $\max_{k\neq j}|(Z_j^0)^T \bx_k|/n =
O_P(\sqrt{\log(p)/n})$ (see Lemma \ref{lemm1} and using (C1,i) and (C1,ii)): 
\begin{eqnarray}\label{add3}
\|\bx_{-j}(\hat{\gamma}_j - \gamma_j^0) \|_2^2/n = o_P(1).
\end{eqnarray}
By (\ref{add1a}), (\ref{add2}) and (\ref{add3}) we complete the
proof.\hfill$\Box$

\bibliographystyle{apalike} 
\bibliography{reference}

\end{document}